\documentclass[10pt]{article}
\input epsf
\def\kb {k_{{B}}}

\begin{document}

\title{\bf Brillouin scattering and the CMB}

\author{A. Sandoval-Villalbazo$^a$ and R. Maartens $^{b}$\\
$^a$ Departmento de F\'{\i}sica y Matem\'{a}ticas,
Universidad Iberoamericana\\
Lomas de Santa Fe 01210 M\'{e}xico D.F., M\'{e}xico \\
$^b$ Institute of Cosmology and Gravitation \\
University of Portsmouth, Portsmouth PO1 2EG, UK \\
E-mail: alfredo.sandoval@uia.mx \\
E-mail: roy.maartens@port.ac.uk}

\maketitle
\bigskip

\begin{abstract}

Brillouin scattering of photons off the density fluctuations in a
fluid is potentially important for cosmology. We derive the
Brillouin spectral distortion of blackbody radiation, and discuss
the possible implications for the cosmic microwave background. The
thermal Sunyaev-Zeldovich effect is slightly modified by Brillouin
distortion, but only at very long wavelengths.

\end{abstract}

\vskip 1cm

\section{Introduction}

Dynamical light scattering theory~\cite{Berne} shows that photons
interact with statistical fluctuations in the fluid density and
temperature. The origin of these fluctuations can be traced back
to the fact that classical particles possess random trajectories
that allow a description in terms of a dynamic structure factor.
Early theoretical work by Brillouin~\cite{Brillouin} predicted a
doublet in the frequency distribution of monochromatic light
scattered by sound waves in a fluid. Experimental
work~\cite{Gross} soon confirmed the Brillouin doublet about the
central or Rayleigh line. Landau and Placzek~\cite{Landau1} gave a
theoretical explanation of the Brillouin spectrum, by means of a
thermodynamical approach. Brillouin scattering has been mainly
investigated in the context of laboratory physics. We are not
aware of applications to cosmology, although Brillouin scattering
has been considered in planetary atmospheres~\cite{kyh} and pulsar
eclipses~\cite{lc}.

In this paper, we derive the Brillouin spectral distortion of
blackbody radiation interacting with a hydrodynamical fluid. We
discuss the possible applications of this result to the cosmic
microwave background (CMB) radiation. We also investigate the
interaction of the CMB with hot intra-cluster gas. The thermal
Sunyaev-Zeldovich (SZ) effect is slightly modified by Brillouin
scattering, but only at very long wavelengths: the CMB is
``protected" from thermal Brillouin distortion by the low density
and high temperature of the intra-cluster gas, which means that it
does not behave like a hydrodynamical fluid.

\section{The Brillouin spectrum}

The standard system describing fluctuations of the thermodynamical
variables in a hydrodynamical fluid of particles of mass $m$ can
be written as~\cite{Berne}:
\begin{eqnarray}
\frac{\partial }{\partial t}\,\delta n +n_{o}\,\delta \Theta
&=&0\,,  \label{Nav1} \\ \frac{\partial }{\partial t}\, \delta
\Theta  +\frac{c_{\rm s}^{2}}{ n_{o}}\,\nabla ^{2}\, \delta n +
\alpha c_{\rm s}^{2} \, \nabla ^{2}\, \delta T &=&0\,,
\label{Nav2} \\
\frac{\partial }{\partial t}\, \delta T +\left( \frac{1-\gamma }{
\alpha n_{o}}\right) \frac{\partial }{\partial t}\, \delta n
&=&0\,,  \label{Nav3}
\end{eqnarray}
where $\delta n$ is the fluctuation of the particle density
relative to its average value $n_{o}$, $\Theta=\nabla \cdot
\vec{v} $ is the fluid expansion rate, $c_{\rm s}^{2}$ is the
sound speed, $\alpha =-n^{-1}({\partial n}/{\partial T} )_{p}$ is
the thermal expansion coefficient, $\delta T$ is the temperature
fluctuation, and $\gamma =C_{p}/C_{V}$ is the heat capacity ratio.

Following Landau and Placzek~\cite{Landau1}, a spatial Fourier
transform is taken of Eqs.~(\ref{Nav1})--(\ref{Nav3}) and then a
Laplace transform is performed with respect to time:
\begin{eqnarray}
\left[
\begin{array}{ccc}
s & n_{o} & 0 \\ -c_{\rm s}^{2} k^{2}/(\gamma n_{o}) & s & -\alpha
c_{\rm s}^{2} k^{2}/\gamma \\ (1- \gamma) s/({\alpha n_{o}}) & 0 &
s
\end{array}
\right] \left[
\begin{array}{c}
\delta \tilde{n}\left( k,s\right) \\ \delta \tilde{\Theta}\left(
k,s\right) \\ \delta \tilde{T}\left( k,s\right)
\end{array}
\right] = \left[
\begin{array}{c}
\delta \tilde{n}\left( k,0\right) \\ \delta \tilde{\Theta}\left(
k,0\right) \\ \delta \tilde{T}\left( k,0\right) + (1-
\gamma)\delta \tilde{n}\left( k,0\right) /({\alpha n_{o}})
\end{array}
\right].  \label{sistem1}
\end{eqnarray}
The solution for the density fluctuations in Fourier space is
\begin{equation}
\delta \tilde{n}\left( k,t\right) = \delta \tilde{n}\left(
k,0\right) \left[ \left( 1-\frac{1}{\gamma }\right)
+\frac{1}{\gamma } \cos(c_{\rm s} kt)\right]\,. \label{NDF}
\end{equation}

The Brillouin specific intensity of the scattered light due to its
interaction with acoustic modes of a fluid is given by
\begin{equation}
I_\omega\propto \int_{-\infty}^\infty \langle  \delta
\tilde{n}\left( k,0\right)^* \delta \tilde{n}\left( k,t\right)
\rangle \, {\rm e}^{-i\omega t}\,dt\,,
\end{equation}
which leads to
\begin{equation}
I_{\nu }\propto \left(1-\frac{1}{\gamma }\right)\delta (\nu
)\,+\frac{1}{2 \gamma } \delta (\nu +\frac{c_{\rm s}}{c}
\nu)+\frac{1}{2 \gamma }\delta (\nu -\frac{c_{\rm s}}{c} \nu).
\label{Intensity1}
\end{equation}
The Brillouin spectrum is a sum of Dirac delta functions. In the
presence of dissipative effects, the delta peaks acquire finite
height and width, but we are neglecting non-equilibrium effects,
which would typically (but not always) produce only a small
correction to our equilibrium results. The first term in
Eq.~(\ref{Intensity1}) represents the Rayleigh peak. The next two
terms are the Brillouin doublet. This doublet reflects a shift in
frequency governed by the speed of sound of the fluid, directly
related to the Doppler effect, as shown by Eq.~(\ref{NDF}):
\begin{equation}
\Delta \nu =\frac{c_{\rm s}}{c} \nu \,.  \label{Shift}
\end{equation}
The adiabatic speed of sound for a monatomic gas is given by
\begin{equation}
c_{\rm s}={\frac{\kb T}{m }\gamma }\,.  \label{sound}
\end{equation}

The integrated intensities under each peak are simply related by
the Landau-Placzek ratio. If we denote by $I_{{\rm s}}$ the area
under the singlet peak, and $I_{{\rm d}}$ the area under one of
the doublet peaks, then
\begin{equation}
I=\int_{-\infty }^{\infty }I_{\omega }\,d\omega =I_{{\rm
s}}+2I_{{\rm d}}\,, \label{Ratio1}
\end{equation}
where
\begin{equation}
\frac{I_{{\rm s}}}{I_{{\rm d}}}=\gamma -1\,.  \label{LPR}
\end{equation}
Thus, for a monatomic gas with $\gamma =\frac{5}{3}$, it follows
that $\frac{ 1}{4}$ of the total intensity of light scattered from
a fixed frequency $ \nu $ will remain unshifted, while
${\frac{3}{4}}$ of the intensity will be shifted in equal parts to
the frequencies $\nu \pm \Delta \nu $, given by Eq.~(\ref{Shift}).
Also, if the density of the fluid increases, $\gamma \rightarrow
1$ and the Rayleigh peak disappears, leaving all the scattered
radiation divided by equal parts in the Brillouin doublet.

In the hydrodynamical limit, density fluctuations propagate as
sound waves and cause the doublet. In this limit, sound waves of
all wavenumbers are possible, so that all photon frequencies are
affected by Brillouin scattering. We will discuss later how this
is changed in the non-hydrodynamical regime.

The undistorted blackbody spectrum is
\begin{equation}
I=2 {(\kb T_{r} )^{3}\over(h c)^2}F(x)\,,
\end{equation}
where
\begin{equation}
x=\frac{h \nu }{\kb T_{r}} \,,~~~ F(x) ={\frac{x^{3}}{e^{x}-1}}\,,
\end{equation}
and $T_{r}$ is the radiation temperature. The effect of Brillouin
scattering on the intensity spectrum can be calculated from
Eqs.~(\ref{Intensity1})--(\ref{LPR}), for the case $\gamma
\rightarrow 1$, by means of the convolution integral ~\cite{JPA}:
\begin{equation}
I_{\rm br}\left( \nu \right) =[1-\tau_{\rm br}]I(
\nu)+\frac{\tau_{\rm br}}{2}\int_{-\infty}^{\infty }I( \bar{\nu})
\left\{\delta \left(\bar{\nu}-\nu \left[1+\frac{c_{\rm
s}}{c}\right]\right)+\delta \left(\bar{\nu}-\nu
\left[1-\frac{c_{\rm s}}{c}\right]\right) \right\} \,d\bar{\nu}
\,. \label{conv}
\end{equation}
Here $\tau_{\rm br}$ is the fraction of photons that are Brillouin
scattered off the acoustic modes of the fluid and $I_{\rm
br}=I+\Delta I_{\rm br} $ is the perturbed spectrum. For a fairly
dense system, $\tau_{\rm br} \sim 1$.

The Brillouin peaks are Dirac delta functions, preserving the
intensity fractions of the scattered radiation. If we now perform
the substitution
\begin{equation}
\alpha=\bar{\nu}-\nu \left(1\pm \frac{c_{\rm s}}{c}\right) \,,
\label{alfa1}
\end{equation}
we may expand the integrand in a Taylor series. Straightforward
calculation leads to the following expression for CMB scattering
off the acoustic modes:
\begin{equation}
{\Delta I_{\rm br}}= \frac{{\tau_{\rm br}}}{2} \left( \frac{c_{\rm
s}}{c}\right)^{2} \nu^{2}\, \frac{\partial^{2} I}{\partial
\nu^{2}}\,. \label{alfa2}
\end{equation}
This expression is plotted in Fig.~1 for $c_{\rm s}=5.7 \times
10^{7}~{\rm cm/s}$, which roughly corresponds to a fluid
temperature of 4000~K. The curve resembles the ordinary thermal SZ
effect, with the "crossover frequency" in this case being
$\nu_{\rm a,c}=262.612~{\rm GHz}$.

\begin{figure}
\epsfxsize=3.4in \epsfysize=2.6in
\epsffile{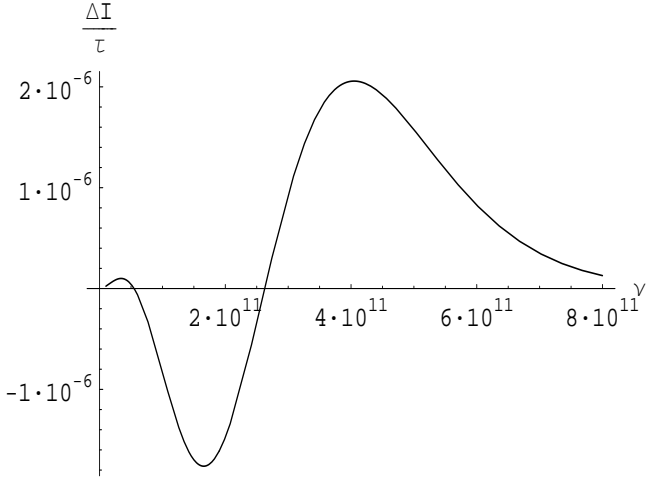}\vspace{0.5cm} \caption {CMB distortion for
the case of Brillouin scattering off hydrodynamical acoustic
modes, as given by Eq.~(\ref{alfa2}). The fluid temperature is
$T=10^{4}~{\rm K}$. The frequency is given in Hz and ${\Delta
I_{\rm br}}/{\tau_{\rm br}}$ is given in units of ${2 (k
T_{r})^{3}}/{(h c)^{2}}$. }
\end{figure}
\vspace{0.5cm}

\section{Corrections to the SZ effect?}

Consider now the case of CMB photons crossing the intra-cluster
gas of a galaxy cluster. The Brillouin distortion is apparently
given by Eq.~(\ref{alfa2}), which may be compared with the
(non-relativistic, thermal) SZ distortion~\cite{ZS1}
\begin{equation}\label{sz}
\Delta I_{{\rm sz}}=yI\,\frac{e^{x}F(x)}{2x^{2}} \left[ x\coth
\frac{x}{2}-4\right] \,.
\end{equation}
However, the intra-cluster gas is hot ($T\sim 10^8$~K) and very
dilute ($n\sim 10^3$~m$^{-3}$). This means that it is far from the
hydrodynamic (collision-dominated) regime, and much closer to the
collisionless regime. Therefore the hydrodynamical analysis above
does not cover the intra-cluster gas. Pressure fluctuations do not
disappear, but they no longer propagate as sound waves at all
frequencies. While there may be sound-like waves, there will be a
maximum frequency, above which there is no propagation~\cite{i}.

An interesting question is how the transition from propagating to
non-propagating frequencies is determined, but answering this
question requires a detailed kinetic-theory analysis, which will
be tackled elsewhere. For present purposes, we can assume that
there is an instantaneous cut-off at about the plasma frequency.
For typical clusters, this gives an upper limit
\begin{equation}\label{lim}
k \leq  k_{\rm max}\sim k_{\rm plasma} \sim
10^{-4}~\mbox{m}^{-1}\,.
\end{equation}
This means that standard Brillouin scattering of the CMB by
clusters is cosmologically insignificant, since it affects only
very low frequencies.

Nevertheless it is interesting to compute the distortion up to the
cut-off frequency, which is given by Eq.~(\ref{Intensity1}),
subject to Eq.~(\ref{lim}), in the case $\gamma=\frac{5}{3}$. The
corresponding result now reads:
\begin{equation}\label{brill3}
\Delta I=\frac{1}{4 \gamma} \left(\frac{c_{\rm s}}{c}\right)^{2}
\nu^{2}\, \frac{\partial^{2} I}{\partial \nu^{2}}\,.
\end{equation}

The Brillouin distortion of the CMB by a typical cluster over the
allowable range of frequencies is shown in Fig.~2. The SZ
distortion in this range is also shown, for comparison. Although
the Brillouin contribution is not negligible, it operates at
frequencies which are of little interest.

\begin{figure}
\epsfxsize=3.4in \epsfysize=2.6in
\epsffile{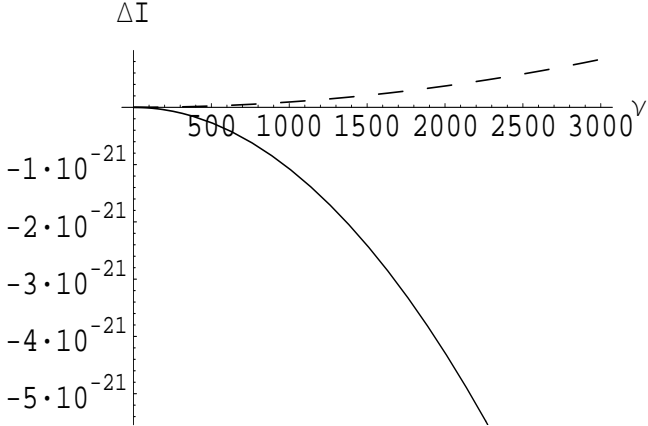}\vspace{0.5cm} \caption {The solid line
represents the ordinary SZ effect in the given frequency range.
The dashed line corresponds to Brillouin scattering. The cluster
parameters are $T=10^{7}~{\rm K}$, and optical depth
$\tau=10^{-3}$. The units are as in Fig.~1.
 }
\end{figure}
\vspace{0.5cm}

\section{Discussion}

We derived the distortion of a blackbody spectrum that follows
from the hydrodynamical Brillouin frequency shift, given in
Eq.~(\ref{alfa2}) and illustrated in Fig.~1. The question then is
what are the cosmological implications of this result. We
considered the Brillouin distortion induced by clusters, but since
the intra-cluster gas is not hydrodynamical, the standard analysis
does not apply beyond a very low frequency, so that Brillouin
distortion is insignificant.

The ordinary Brillouin spectrum arises from light scattering off
pressure fluctuations at constant entropy. When the fluid is
collision-dominated, these fluctuations propagate as sound waves
and generate a Brillouin doublet for each wavenumber. As density
decreases (and the Knudsen number approaches 1), the concept of
sound propagation eventually becomes meaningless. For clusters,
there is a cut-off maximum wavenumber for propagating modes, of
roughly the plasma frequency, i.e. $\sim 10^{-4}~{\rm m}^{-1}$.
Thus the CMB is ``protected" from Brillouin distortion over all
frequencies except the very low tail. The SZ effect is therefore
in practice unaffected by ordinary Brillouin scattering.

Nevertheless, there may still be some Brillouin-type affects on
the CMB from clusters. Pressure fluctuations still exist beyond
the hydrodynamic regime and, although no doublet is present in the
scattered spectrum, each central (Rayleigh) line will be broadened
by an quantity proportional to the ``width parameter"~\cite{gb}
\begin{equation}
s(\nu)={2\pi\nu\over c}\left({\kb T\over m}\right)^{1/2}\,.
\end{equation}
This means that some scattering of photons will occur due to
pressure fluctuations even in the collisionless limit. The
distortion induced by this broadening is of the form given by the
convolution integral (compare~\cite{sfyf})
\begin{equation}
I+\Delta I \propto \int_0^\infty{I(\nu')\over
s(\nu')}\exp-\left[{\nu'- \nu\over s(\nu')}\right]^2d\nu'\,.
\end{equation}
This non-standard collisionless Brillouin distortion could lead to
effects at significant frequencies, and this is currently under
investigation.

A Boltzmann equation approach to the study of time correlation
functions shows that sound-wave peaks are present for wavelengths
comparable to the mean free path~\cite{yn}. Furthermore, because
of the high densities of hot photons in the intra-cluster gas,
there may be propagating modes at higher frequencies; relativistic
plasmas at high temperatures ($T>10^8$~K) may in some sense be
treated as collision-dominated systems~\cite{glw}. Other arguments
that suggest collective behaviour of low density plasmas far
beyond the plasma frequency have been developed in~\cite{bbp}: the
plasma-dynamical regime describes motions of spatially smooth,
weakly damped disturbances which may oscillate at high
frequencies.

In view of these points, we believe that dynamical light
scattering of CMB photons in clusters should not be simply
dismissed, and further investigation is warranted. Furthermore,
there remains the potentially more important question of whether
ordinary Brillouin scattering has an effect during recombination.

\[ \]
{\bf Acknowledgements:}

We thank Naoki Itoh for pointing out the limits in which Brillouin
scattering operates. We also thank Leopoldo Garc\'{i}a-Col\'{i}n,
Bruce Bassett, Qinghuan Luo and Thanu Padmanabhan for valuable
comments. ASV was supported during this work by  CONACyT grant
41081-F, and thanks the Institute of Cosmology and Gravitation at
Portsmouth for hosting him while this work was done. RM thanks the
ESI Mathematical Cosmology Programme, Vienna, where part of this
work was completed. The work of RM is supported by PPARC.

\end{document}